\documentclass{article}
\usepackage{makeidx}
\usepackage{amsfonts}
\usepackage{amssymb}
\usepackage{amsmath}

\newtheorem{lemma}{Lemma}
\newtheorem{proposition}{Proposition}

\begin{document}
\title{The entropy gain of infinite-dimensional quantum channels}
\author{A. S. Holevo\\ Steklov Mathematical Institute}
\date{}
\maketitle
\begin{abstract}
In the present paper we study  the entropy gain $H(\Phi [\rho])-H(\rho )$ for infinite-dimensional channels $\Phi$. We show that unlike finite-dimensional case where the minimal entropy gain is always nonpositive  \cite{al}, there is a plenty of channels with positive minimal entropy gain. We obtain the new lower bound and compute the minimal entropy gain for a broad class of Bosonic Gaussian channels by proving that the infimum is attained on the Gaussian states.
\end{abstract}
\section{Introduction}
For a channel $\Phi$ and an input state $\rho$ the \textit{entropy gain} is given by $H(\Phi [\rho])-H(\rho )$, where $H(\rho )=-\mathrm{Tr}\rho \log\rho$ is the von Neumann entropy\footnote{Throughout the paper $\log$ will denote the natural logarithm.} of the density operator $\rho$ in the system Hilbert space $\mathcal{H}$.
In the paper \cite{al} the minimal entropy gain $G(\Phi)=\inf_\rho(H(\Phi [\rho])-H(\rho))$ was studied for the case $\dim \mathcal{H}=d<\infty$, and it was shown that
\begin{enumerate}
  \item  $  -\log d\leq G(\Phi)\leq 0;$
  \item $G(\Phi)$ is additive w.r.t. tensor product of channels.
\end{enumerate}

The inequality $G(\Phi)\leq 0$ was derived in \cite{al} from the observation that in finite dimensions every channel has an invariant state.
This follows also directly from the estimate $\inf_\rho(H(\Phi [\rho])-H(\rho))\le \inf_\rho(\log d-H(\rho))=0$.
It implies, in particular, that $G(\Phi)= 0$ for finite-dimensional unital channels since the entropy gain is always nonnegative in this case \cite{wehrl}.

In the present paper we study similar quantity for infinite-dimensional channels. First of all, the entropy gain needs to be correctly defined since the entropies can assume the value $+\infty$. Then the statement 1. concerning the possible range of $G(\Phi)$ changes substantially and there is a plenty of channels with positive minimal entropy gain. We obtain the new lower bound and compute the minimal entropy gain for a broad class of Bosonic Gaussian channels by proving that the infimum is attained on the Gaussian states. The additivity property still holds and its proof (which is a simple corollary of the strong subadditivity) does not change provided the entropy gains are well defined.

\section{Lower bound for the entropy gain}\label{LB1} If $\{A_n\}$ is a monotone sequence of positive operators converging weakly (and hence strongly) to the bounded operator $A$, we write $A_n\uparrow A$.

Let $\Phi$ be a channel in separable Hilbert space $\mathcal{H}$ defined by the Kraus representation
\begin{equation}\label{1}
\Phi [\rho]=\sum_{j=1}^{\infty}V_j\rho V_j^*;\quad \sum_{j=1}^{\infty}V_j^*V_j=I,
\end{equation}
where $\rho$ is a density and hence trace class operator. Since  the unit operator  $I$ in the infinite dimensional case is not trace class, the expression $\Phi[I]$ in general is not defined. However there is an important case when it still can be naturally defined as a bounded positive operator. Consider the condition
 \begin{equation}\label{0}
\sum_{j=1}^{\infty}\|V_j^*\psi\|^2<\infty,\quad\psi\in \mathcal{H},
\end{equation}
which, by the uniform boundedness principle \cite{reed}, implies $\sum_{j=1}^{n}V_jV_j^*\uparrow A$, where $A$ is positive bounded operator which we denote $\Phi[I]$. Note that by the relation between different Kraus representations this operator does not depend on the choice of representation. Such channels will be called \textit{regular} in this paper. An example of channel which is not regular is given by the completely depolarizing channel $\Phi [\rho]=\rho_0\mathrm{Tr}\rho$ for trace-class operators $\rho$.

In the following we need an extension of the quantum expectation functional to unbounded operators.
Let $F$ be a selfadjoint positive operator. For any density operator $\rho$ we define
\begin{equation}\label{M}
\mathrm{Tr}\rho F=\sum_{k=1}^{\infty}\lambda _{k}\langle e_k|F | e_k\rangle ,
\end{equation}
where $\lambda _{k}$ are the eigenvalues and $e_k$ are the eigenvectors of $\rho$, having in mind that $\langle e_k|F |e_k\rangle= +\infty$ if $e_k\not\in \mathcal{D}(\sqrt{F}).$

\begin{lemma} Let $E(d\lambda)$ be the spectral measure of the operator $F$, and $m_\rho (B)=\mathrm{Tr}\rho E(B)$ for Borel $B\subset\mathbb{R}_+$, then

\begin{equation*}
\mathrm{Tr}\rho F=\int_0^{\infty}\lambda m_\rho (d\lambda ).
\end{equation*}\end{lemma}

In general, $0\le\mathrm{Tr}\rho F\le +\infty$. This can be further generalized to positive operators $F$ ``which may assume the value $+\infty$'' if we define them via a spectral measure on the extended real half-line $\overline{\mathbb{R}}_+=[0,+\infty]$. For brevity we denote the class of such operators $\mathfrak{F}(\overline{\mathbb{R}}_+)$. For given $F$ we denote $\mathfrak{S}_F(\mathcal{H})=\{\rho: \mathrm{Tr}\rho F<\infty\}$. The whole construction then obviously extends to selfadjoint operators bounded from below.

For a density operator $\sigma$ we have $-\log\sigma\in \mathfrak{F}(\overline{\mathbb{R}}_+)$.
The relative entropy is defined as
\begin{equation*}
H(\rho ||\sigma)=\sum_{k=1}^{\infty}\lambda _{k}[\log\lambda _{k}-\langle e_k|\log \sigma| e_k\rangle ]
\end{equation*}
and in any case we have
\begin{equation}\label{F}
H(\rho ||\sigma)+H(\rho)=\mathrm{Tr}\rho (-\log \sigma),
\end{equation}
where both sides are nonnegative but can be infinite.

If the positive selfadjoint operator $F$ is such that
\begin{equation}\label{BF}
\mathrm{Tr}\exp (-\beta F)<\infty,\quad \beta >0,
\end{equation}
then the formula
\begin{equation}
\rho _{\beta }=\exp (-\beta F-c(\beta ));\quad c(\beta )=\log \mathrm{Tr}\exp
(-\beta F),  \label{2}
\end{equation}
defines a density operator. Then usually $F$ has meaning of a Hamiltonian and $\rho _{\beta }$ is the Gibbs (thermal) state at inverse
temperature $\beta$ \cite{wehrl}. In this case $\mathfrak{S}_F(\mathcal{H})=\{\rho: \mathrm{Tr}\rho F<+\infty\}$ is the subset of states with finite energy.
When $\beta\to 0$, one has $\exp (-\beta F)\uparrow I$ , and for a regular channel, $\Phi[\exp (-\beta F)]\uparrow \Phi[I]$.

Let $\rho$ be any density operator with finite entropy $H(\rho)=\sum_{k=1}^{\infty}\lambda_k(-\log\lambda _{k})$. Then an operator $F$ satisfying $\mathrm{Tr}\rho F<+\infty$ and (\ref{BF}) always exists. Indeed, take
$$F=\sum_{k=1}^{\infty}\mu_k(-\log\lambda _{k}) |e_k\rangle \langle e_k |,$$
where $\mu_k\uparrow+\infty$  such that $\sum_{k=1}^{\infty}\mu_k\lambda_k(-\log\lambda _{k})$ still converges\footnote{For this remark the author is indebted to M. E. Shirokov, who used a similar trick in his paper \cite{sh}.}.

\begin{proposition}\label{P1} Let $\Phi$ be a regular channel and $\rho$ a state with finite entropy, then
\begin{equation}\label{3}
H(\Phi [\rho])-H(\rho)\ge \mathrm{Tr}\Phi [\rho](-\log\Phi[I]),
\end{equation}
where on the right stands the extended quantum expectation of the operator $-\log\Phi[I]$ which is selfadjoint and bounded from below.
\end{proposition}

\textit{Proof.} Since $H(\rho)<\infty$, the entropy gain is unambiguously defined as a quantity with values in $(-\infty,+\infty].$ By the remark above we can choose $F$ satisfying (\ref{BF}) such that $\rho\in\mathfrak{S}_F(\mathcal{H})$. From (\ref{BF}), (\ref{2}) it follows that
\begin{equation}\label{star}
H(\rho ||\rho _{\beta })+H(\rho)=\beta\mathrm{Tr}\rho F +c(\beta ),
\end{equation}
where all terms are finite.
By monotonicity of the relative entropy
\begin{equation}
H(\Phi[\rho] ||\Phi[\rho _{\beta }])\leq H(\rho ||\rho _{\beta }),
\label{4}
\end{equation}
whence by (\ref{F})
\begin{equation*}
\mathrm{Tr}\Phi [\rho](-\log\Phi[\rho _{\beta }])\leq H(\Phi[\rho]) - H(\rho)+\beta\mathrm{Tr}\rho F +c(\beta ).
\end{equation*}
After insertion of (\ref{2}) the term $c(\beta )$ cancels and we obtain
\begin{equation}
\mathrm{Tr}\Phi [\rho](-\log\Phi[\exp (-\beta F)])-\beta\mathrm{Tr}\rho F\leq H(\Phi[\rho])- H(\rho),
\label{5}
\end{equation}
where $-\log\Phi[\exp (-\beta F)]$ is selfadjoint operator bounded from below.
Since $\exp (-\beta F)\le I$, we have  $\log\Phi[\exp (-\beta F)]\le\log\Phi[I]$ by the operator monotonicity
of the function $\log x$ on $\mathbb{R}_+$.
Letting $\beta\to 0$, equation (\ref{5}) implies (\ref{3}).$\square$

Thus the definition of the minimal entropy gain  should be modified as
\begin{equation}\label{MG}
G(\Phi)=\inf_{\rho: H(\rho)<\infty}\left(H(\Phi [\rho])-H(\rho)\right).
\end{equation}
Taking into account that $0\le\Phi[I]\le\|\Phi[I]\|I$ we have $-\log\Phi[I]\ge (-\log\|\Phi[I]\|)I$ and thus
\begin{equation}\label{LB}
-\log\|\Phi[I]\|\le G(\Phi).
\end{equation}

In particular, defining \textit{unital} channels as regular channels with $\Phi[I]=I$, we have $G(\Phi)\ge 0$.
In infinite dimensions there are even classical unital channels with $G(\Phi)=+\infty$. To show this consider the infinite stochastic
matrix $\Phi=[p_{ij}]_{i,j\in\mathbb{N}_+}$, where  $p_{ij}=q_{n_j(i)},\,Q=\{q_n\}$ is a probability distribution on $\mathbb{N}_+=\{1,2,\dots\}$ with infinite entropy and $n_j(i)$ are the permutations of $\mathbb{N}_+$ defined in the Appendix, lemma \ref{per}. Then the matrix is doubly stochastic, i.e. $\Phi$ is unital and for any pure classical state $\delta_i$ we have $H(\Phi [\delta_i ])=H(Q)=+\infty$. Hence, by concavity of the entropy, $H(\Phi [P])=+\infty$ for any probability distribution $P$ on $\mathbb{N}_+$.

\section{The case of Bosonic Gaussian channels}\label{GC}
Let $(Z, \Delta )$ be a coordinate symplectic space ($\dim Z=2s$) with the nondegenerate skew-symmetric commutation matrix $\Delta$, and let $W(z)=\exp (i R z);\,z\in Z$ be the Weyl system in a Hilbert space $\mathcal{H}$ giving the representation
for the Canonical Commutation Relations. Here $R$ is the $2s$-vector row of the canonical variables.
Gaussian state $\rho$ with zero mean and the real positive definite  covariance matrix $\alpha$ satisfying $\alpha\pm \frac{i}{2}\Delta\ge 0$ is defined by the characteristic function
\begin{equation}\label{CF}
\phi_\rho (z)=\mathrm{Tr}\rho W(z)=\exp \left(-\frac{1}{2}z^{\top }\alpha z\right).
\end{equation}
The state is nondegenerate if and only if
\begin{equation}\label{ND}
\alpha -\frac{i}{2}\Delta >0,
\end{equation}
i.e. the complex positive semidefinite matrix $\alpha -\frac{i}{2}\Delta$ is nondegenerate (and thus positive definite, see Appendix).

Let $\Phi $ be a (centered) Bosonic Gaussian channel,
\begin{equation*}
\Phi ^{\ast }\left( W(z)\right) =W(Kz)\exp \left( -\frac{1}{2}
z^{\top }\mu z\right) ,
\end{equation*}
where the real positive semidefinite matrix $\mu$ satisfies
\begin{equation}
\mu \geq  \pm \frac{i}{2}\left( \Delta
-K^{\top }\Delta K\right) .  \label{nid}
\end{equation}
The channel $\Phi$ transforms a Gaussian state with covariance matrix $\alpha$ into Gaussian state with covariance matrix
\begin{equation}\label{CM}
\alpha'=K^\top\alpha K +\mu.
\end{equation}See \cite{h}, \cite{h2}, \cite{qi} for more detailed descriptions.

We take $F=R\epsilon R^{\top },$ where $\epsilon$ is a nondegenerate positive definite matrix (e.g. unit matrix).
Notice that with such choice $\mathfrak{S}_F(\mathcal{H})$ coincides with the subset of states  with finite second moments.

\begin{lemma} The density operator $\rho _{\beta}$ given by (\ref{2}) is
the density operator of nondegenerate Gaussian state with zero mean and covariance matrix $\alpha_\beta$ satisfying
\begin{equation}
2\Delta ^{-1}\alpha_\beta =\cot \beta\epsilon \Delta.  \label{cot1}
\end{equation}
Moreover,
\begin{equation}
c(\beta )=\frac{1}{2}\log\left[\det \left( \Delta ^{-1}\alpha_\beta -\frac{i}{2}I\right) \right].
\label{constan}
\end{equation}
and the entropy is
\begin{equation}
H(\rho _{\beta})=\frac{1}{2}\log\det \left[\Delta ^{-1}\alpha_\beta
-\frac{i}{2}I \right] +\mathrm{Sp}(\Delta ^{-1}\alpha_\beta
)\mathrm{arc}\cot \left( 2\Delta
^{-1}\alpha _\beta\right)  \label{entr},
\end{equation}
where $\mathrm{Sp}$ denotes trace of $2s\times 2s$-matrix.\end{lemma}

The proof of this lemma is given in the Appendix.

\begin{proposition}\label{P2} Let the matrix $K$ be nondegenerate, then $\Phi$ is regular with
\begin{equation}\label{PI}
\Phi[I]=|\det K|^{-1}I.
\end{equation}
Moreover
\begin{equation}\label{INF}
G(\Phi )=\log |\det K|.
\end{equation}
\end{proposition}

\textbf{Remarks} 1. Note that $\det \Delta =1$ in the case of the canonical form
\begin{equation}
\Delta =\left[
\begin{array}{ccccc}
0 & -1 &  &  &  \\
1 & 0 &  &  &  \\
&  & \ddots &  &  \\
&  &  & 0 & -1 \\
&  &  & 1 & 0
\end{array}
\right] \equiv \mathrm{diag}\left[
\begin{array}{cc}
0 & -1 \\
1 & 0
\end{array}
\right] .  \label{delta}
\end{equation}

2. In the case of one mode ($s=1$) there are three examples of special interest: attenuator with coefficient $k<1$, amplifier with $k>1$ and the channel with additive classical Gaussian noise ($k=1$), see \cite{h2}. In all these cases the minimal entropy gain is equal to $\log k^2$
giving all possible real values.

\textit{Proof.} By using (\ref{constan}) we have
\begin{equation}\label{J}
\Phi [\exp (-\beta F)]=\left[\det \left( \Delta ^{-1}\alpha_\beta -\frac{i}{2}I\right)\right]^{1/2}\Phi [\rho _{\beta}].
\end{equation}
The state $\Phi [\rho _{\beta}]$ has the covariance matrix $\alpha'_\beta = K^\top\alpha_\beta K +\mu$, hence by the inversion formula
for characteristic functions
\begin{equation}\label{I}
\Phi [\rho _{\beta}]=\frac{1}{(2\pi )^{s}}\int\exp\left(-\frac{1}{2}z^\top \alpha'_\beta  z\right)W(-z)d_\Delta^{2s}z,
\end{equation}
where $d_\Delta^{2s}z=\sqrt{\det \Delta}\,d^{2s}z$ is the element of the symplectic volume, corresponding to $\Delta$.

Now consider the asymptotic $\beta\to 0$. Then from (\ref{cot1})
\begin{equation}\label{asym}
\alpha_\beta\sim\frac{1}{2\beta}\epsilon^{-1}
\end{equation}
and hence
\begin{equation}\label{K}
\det\alpha'_\beta \sim |\det K|^2\det \alpha_\beta \sim |\det K|^2 \det \Delta\det\left( \Delta ^{-1}\alpha_\beta -\frac{i}{2}I\right).
\end{equation}
The probability measure
$$
(2\pi)^{-s}\left(\det\alpha'_\beta \right)^{1/2}\exp\left(-\frac{1}{2}z^\top\alpha'_\beta z\right)d^{2s}z
$$
converges weakly to the probability measure degenerated at $0$ when $\beta\to 0$, therefore  summarizing (\ref{J}), (\ref{I}), (\ref{K}), one obtains that the operators
$\Phi [\exp \left( -\beta F\right)]$ converge to $|\det K|^{-1}W(0)$ in the weak operator topology. Taking into account that $W(0)=I$ we obtain (\ref{PI}) and proposition \ref{P1} implies
\begin{equation}\label{33}
H(\Phi [\rho])-H(\rho)\ge \log |\det K|
\end{equation}
for any state $\rho$ with finite entropy.

Let us show that the bound (\ref{33}) is achieved asymptotically for the states $\rho _{\beta},\,\beta\to 0$. Equations (\ref{entr}), (\ref{asym}), (\ref{K}) imply that the main term in (\ref{entr}) is the first one giving
\begin{equation}\label{asent}
H(\rho _{\beta})\sim \frac{1}{2}\log\det\left( \Delta ^{-1}\alpha_\beta -\frac{i}{2}I\right)\sim \frac{1}{2}\log\frac{\det \alpha_\beta}{\det \Delta}
\end{equation}
and similarly
\begin{equation}\label{asent1}
H(\Phi [\rho _{\beta}])\sim\frac{1}{2}\log\frac{\det \alpha'_\beta}{\det \Delta}
\end{equation}
whence by (\ref{K})
\begin{equation}\label{dif}
H(\Phi [\rho _{\beta}])-H(\rho _{\beta})\sim \frac{1}{2}\log \frac{|\det K|^2\det\alpha_\beta}{\det \Delta}  -\frac{1}{2}\log\frac{\det \alpha_\beta}{\det \Delta} \sim \log |\det K|.
\end{equation}
$\square$

For states with finite second moments one can prove somewhat more. For simplicity we introduce two additional technical restrictions.

\begin{proposition} For a Gaussian channel $\Phi$ satisfying
\begin{equation}\label{N}
\mu > \frac{i}{2}\left( \Delta
-K^{\top }\Delta K\right)
\end{equation}
and a state $\rho$ with finite second moments satisfying the condition (\ref{ND}) one has
\begin{equation*}
H(\Phi [\rho])-H(\rho)\ge H(\Phi [\rho_G])-H(\rho_G),
\end{equation*}
where $\rho_G$ is the Gaussian density operator with the same first and second moments as $\rho$.
\end{proposition}

\textit{Proof.} The condition  (\ref{ND}) implies that  $\rho_G$ is nondegenerate and hence $\rho_G=\rho_{\beta}$ for the choice of the energy matrix $\epsilon$ according to (\ref{cot1}). From equation (\ref{star}) we obtain
\begin{equation}\label{a}
H(\rho ||\rho _{\beta })+H(\rho )=H(\rho _{\beta })+\beta\mathrm{Tr}(\rho -\rho _{\beta })F=H(\rho _{\beta }),
\end{equation}
since the second moments of $\rho$ and $\rho _{\beta }$ coincide. Similarly,
\begin{equation}
H(\Phi [ \rho ]||\Phi [ \rho _{\beta }])+H(\Phi [ \rho ])=H(\Phi [ \rho
_{\beta }])-\mathrm{Tr}(\Phi [\rho] - \Phi [\rho _{\beta}])\log \Phi [ \rho _{\beta }].  \label{41}
\end{equation}
Since $\Phi $ is Gaussian channel, then the covariance matrices of  $\rho$ and $\rho _{\beta }$
are transformed identically and hence the second moments of $\Phi[\rho]$ and $\Phi[\rho _{\beta }]$ also coincide.
Moreover  $\Phi[\rho _{\beta }]$ is a Gaussian state which is nondegenerate by the conditions (\ref{N}), (\ref{ND}) and hence its logarithm is a quadratic form in the canonical variables $R$.
Therefore the last term in (\ref{41}) vanishes and it reduces to
\begin{equation}\label{b}
H(\Phi [ \rho ]||\Phi [ \rho _{\beta }])+H(\Phi [ \rho ])=H(\Phi [ \rho
_{\beta }]).
\end{equation}
Using (\ref{a}), (\ref{b}) and monotonicity of the relative entropy we get
\begin{equation}\label{X}
H(\Phi [ \rho_{\beta }])-H(\rho _{\beta })\le H(\Phi [ \rho ])-H(\rho ).
\end{equation}
$\square$

\section{Closing remarks}

The results of this note indicate that while the entropy increase in the irreversible evolution is traditionally related to the degree of irreversibility, the essentially relevant parameter is the factor by which the phase space volume is changed during the evolution, which can be less, equal or greater than $1$ for different infinite-dimensional evolutions.
The  results in Sec. \ref{LB1} can be generalized to channels with different input and output spaces however in this case the interpretation of the entropy gain and of the quantity $G(\Phi )$ is not so clear. One instance, where it can be interpreted as
``a measure how well the channel $\Phi$ preserves entanglement'' is $G(\tilde{\Phi })$ where $\tilde{\Phi }$ is the complementary channel to $\Phi$,
cf. \cite{djkr}.

\section{Appendix}

\textit{Proof} of Lemma 2. Consider the Gaussian state with zero mean and the covariance matrix $\alpha$. There is a $\Delta-$
symplectic transformation $T$ such that
\begin{equation}
\tilde{\alpha}=T^{\top }\alpha T=\mathrm{diag}\left[
\begin{array}{cc}
\alpha _{j} & 0 \\
0 & \alpha _{j}
\end{array}
\right] ,  \label{one-mode-alpha}
\end{equation}
where $\alpha _{j}\geq \frac{1}{2},\quad j=1,\dots ,s$. Then $\Delta^{-1}\tilde{\alpha}=\mathrm{diag}\left[
\begin{array}{cc}
0 & \alpha _{j} \\
-\alpha _{j} & 0
\end{array}
\right] $ and $\Delta^{-1}\alpha =T\Delta^{-1}\tilde{\alpha}T^{-1}$
is matrix of the operator with eigenvalues $\pm i\alpha _{j}$.
The operator $\rho$ is unitarily equivalent to the
normal modes decomposition

\begin{equation}
\tilde{\rho}_{K}=\bigotimes_{j=1}^{s}\rho ^{(j)},  \label{charct. fct.}
\end{equation}
with $\rho ^{(j)}$ being the elementary one-mode Gaussian density operator

\begin{equation}
\rho ^{(j)}=\frac{1}{\alpha _{j}+\frac{1}{2}}\left( \frac{\alpha _{j}-\frac{
1 }{2}}{\alpha _{j}+\frac{1}{2}}\right) ^{\tilde{n}_{j}},  \label{one-mode}
\end{equation}
where $\tilde{n}_{j}=\frac{1}{2}\left( \tilde{q}_{j}^{2}+\tilde{p}
_{j}^{2}-1\right) $ is the number operator for the $j-$th mode (see
ch. V of \cite{h}). Here the new canonical variables $=[\tilde{q}_{1},\dots,\tilde{p}_{s}]$
are related to the old ones by the formula $\tilde{R}=RT$.

 Since $\alpha -\frac{i}{2}\Delta=\Delta\left( \Delta^{-1}\alpha -\frac{i}{2}I\right) ,$ the condition that $\alpha -\frac{i}{
2 }\Delta$ is nondegenerate is equivalent to $\alpha _{j}>\frac{1}{2}
,\quad j=1,\dots ,s,$ i.e. the decomposition (\ref{charct. fct.}) has no
pure component. Coming back from (\ref{charct. fct.}), (\ref{one-mode}) to
the initial observables $R$ gives
\begin{equation}
\rho=C\exp \left( -R\epsilon R^{\top }\right) ,  \label{Gibbs}
\end{equation}
where
\begin{equation}
C=\prod\limits_{j=1}^{s}\frac{1}{\sqrt{\alpha _{j}^{2}-\frac{1}{4}}}=\frac{1
}{\sqrt{\det \left( \Delta^{-1}\alpha -\frac{i}{2}I\right) }}
\label{consta}
\end{equation}
and $\epsilon $ is found from
\begin{equation}
2\Delta^{-1}\alpha =\cot \epsilon \Delta.  \label{cot}
\end{equation}

The entropy expression (\ref{entr}) follows from the definition by using equations (\ref{2}), (\ref{cot1}), (\ref{constan}).
$\square $

\begin{lemma}\label{per}
There exits a collection of permutations $i\rightarrow n_j(i)$ of $\mathbb{N}_+$ indexed by $j\in\mathbb{N}_+$, such that for any $i\in\mathbb{N}_+$ the sequence $n_j(i); j=1,2,\dots$ is a permutation of $\mathbb{N}_+$.
\end{lemma}

\textit{Sketch of proof.} Consider the matrix $A_\infty= [n_{ij}]_{i,j\in\mathbb{N}_+}$, satisfying
\begin{enumerate}
  \item $n_{i1}=i,\,n_{1j}=j;\quad i,j\in\mathbb{N}_+$;
  \item The matrix $A_\infty$ has the hierarchical structure: for any $k=1,2,\dots$
  \begin{equation*}
  A_\infty=\left[\begin{array}{cc}
                   A_k & \dots \\
                   \dots & \dots
                 \end{array}\right],
  \end{equation*}
  where $A_k$ is $2^k\times 2^k$-matrix of the form
  \begin{equation*}
  A_k=\left[\begin{array}{cc}
                   A_{k-1} & B_{k-1}\\
                   B_{k-1} & A_{k-1}
                 \end{array}\right],
  \end{equation*}
  and
  \begin{equation*}
  B_k=\left[\begin{array}{cc}
                   C_{k-1} & D_{k-1}\\
                   D_{k-1} & C_{k-1}
                 \end{array}\right],
  \end{equation*}
  where $B_k, C_k, D_k$ are $2^k\times 2^k$-matrices of similar structure.
  \end{enumerate}

One can check by inspection taking $k=1,2,\dots,$ that $A_{\infty}$ is uniquely defined by these conditions and by
letting $n_j(i)=n_{ij}$, one obtains the collection of permutations with the required property. $\square$

\textbf{Acknowledgments.} This work was written when the author was visiting Scuola Normale Superiore di
Pisa.  The author is grateful to Rosario Fazio and Vittorio Giovannetti for the invitation and
fruitful discussions and to Maxim Shirokov for valuable remarks.
The work was also partially supported by RFBR
grant 09-01-07066 and the RAS program ``Mathematical Control
Theory''.

\end{document}